\documentclass[sigconf]{acmart}

\AtBeginDocument{%
  }

\copyrightyear{2026}
\acmYear{2026}
\setcopyright{cc}
\setcctype{by}
\acmConference[ICSE '26]{2026 IEEE/ACM 48th International Conference on Software Engineering}{April 12--18, 2026}{Rio de Janeiro, Brazil}
\acmBooktitle{2026 IEEE/ACM 48th International Conference on Software Engineering (ICSE '26), April 12--18, 2026, Rio de Janeiro, Brazil}
\acmPrice{}
\acmDOI{10.1145/3744916.3773217}
\acmISBN{979-8-4007-2025-3/26/04}

\usepackage{listings}
\usepackage{pifont}
\usepackage{tikz}
\usepackage{wrapfig}
\usepackage{multirow}
\usepackage{fontawesome}
\usepackage{enumitem}
\usepackage{threeparttable}

\usepackage[most]{tcolorbox}
\usepackage[normalem]{ulem}
\newcommand{\paratight}[1]{\smallskip\noindent{\bf {#1}}}

\definecolor{codegreen}{rgb}{0,0.6,0}
\definecolor{codegray}{rgb}{0.5,0.5,0.5}
\definecolor{codepurple}{rgb}{0.58,0,0.82}
\definecolor{backcolour}{rgb}{0.95,0.95,0.92}

\lstdefinestyle{mystyle}{
    backgroundcolor=\color{backcolour},   
    commentstyle=\color{codegreen},
    keywordstyle=\color{magenta},
    numberstyle=\tiny\color{codegray},
    stringstyle=\color{codepurple},
    basicstyle=\ttfamily\footnotesize,
    breakatwhitespace=false,         
    breaklines=true,                 
    captionpos=b,                    
    keepspaces=true,                 
    numbers=left,                    
    numbersep=5pt,                  
    showspaces=false,                
    showstringspaces=false,
    showtabs=false,                  
    tabsize=2
}

\newtcolorbox{finding}[1][]{
    boxrule=0.5pt,
    left=1pt,
    right=1pt,
    top=1pt,
    bottom=1pt,
    colback=black!8,
    colframe=black!55,
    notitle,
    enhanced,
    breakable,
}

\newcommand*\emptycirc[1][3pt]{\tikz\draw (0,0) circle (#1);} 
\newcommand*\halfcirc[1][3pt]{%
  \begin{tikzpicture}
  \draw[fill] (0,0)-- (90:#1) arc (90:270:#1) -- cycle ;
  \draw (0,0) circle (#1);
  \end{tikzpicture}}
\newcommand*\fullcirc[1][3pt]{\tikz\fill (0,0) circle (#1);}

\newcommand{\revise}[1]{{\color{black}{#1}}}
\newcommand{\delete}[1]{}

\newcommand{\code}[1]{{\fontfamily{lmtt}\selectfont{#1}}}

\fontfamily{lmtt}\selectfont

\begin{document}

\title{Rethinking the Evaluation of Secure Code Generation}

\author{Shih-Chieh Dai}
\email{shihchieh.dai@utah.edu}
\orcid{0000-0002-5439-3917}
\affiliation{%
  \institution{University of Utah}
  \city{Salt Lake City}
  \state{Utah}
  \country{USA}
}

\author{Jun Xu}
\email{junxzm@cs.utah.edu}
\orcid{0009-0002-8128-5062}
\affiliation{%
  \institution{University of Utah}
  \city{Salt Lake City}
  \state{Utah}
  \country{USA}
}

\author{Guanhong Tao}
\email{g.tao@utah.edu}
\orcid{0000-0002-4701-1327}
\affiliation{%
  \institution{University of Utah}
  \city{Salt Lake City}
  \state{Utah}
  \country{USA}
}

\begin{abstract}

Large language models (LLMs) are widely used in software development. However, the code generated by LLMs often contains vulnerabilities.
Several secure code generation methods have been proposed to address this issue, but their current evaluation schemes leave several concerns unaddressed.
Specifically, most existing studies evaluate security and functional correctness separately, using different datasets. That is, they assess vulnerabilities using security-related code datasets while validating functionality with general code datasets.
In addition, prior research primarily relies on a single static analyzer, CodeQL, to detect vulnerabilities in generated code, which limits the scope of security evaluation.

In this work, we conduct a comprehensive study to systematically assess the improvements introduced by four state-of-the-art secure code generation techniques. Specifically, we apply both security inspection and functionality validation to the same generated code and evaluate these two aspects together.
We also employ three popular static analyzers and two LLMs to identify potential vulnerabilities in the generated code.
Our study reveals that existing techniques often compromise the functionality of generated code to enhance security. Their overall performance remains limited when evaluating security and functionality together. In fact, many techniques even degrade the performance of the base LLM \revise{by more than 50\%}.
Our further inspection reveals that these techniques often either remove vulnerable lines of code entirely or generate ``garbage code'' that is unrelated to the intended task.
Moreover, the commonly used static analyzer CodeQL fails to detect several vulnerabilities, further obscuring the actual security improvements achieved by existing techniques.
Our study serves as a guideline for a more rigorous and comprehensive evaluation of secure code generation performance in future work.

\end{abstract}



\maketitle

\section{Introduction}

Software development is a time-consuming and repetitive task. The rapid evolution of large language models (LLMs) has greatly benefited software developers. Given task requirements described in natural language, LLMs can generate functional and easy-to-adopt code snippets, significantly accelerating the software development process.
Many LLM-based code assistants are already integrated into IDEs, such as Copilot~\cite{copilot} and Cursor~\cite{cursor}.

However, just like human developers, LLMs can also make mistakes when producing code.
One of the major concerns regarding LLM-generated code is its security.
As code snippets generated by LLMs are increasingly incorporated into industrial-level software and systems, it is critical to ensure that LLM-generated code is free of vulnerabilities that could be exploited by attackers.

To address this, a number of techniques have been proposed to improve the security of LLM-generated code, called \textit{secure code generation} methods.
These techniques typically involve either adjusting LLM weight parameters or manipulating the input to or output from the LLM.
For example, SVEN~\cite{he2023large} and SafeCoder~\cite{he2024instruction} construct a training dataset containing code snippets with and without vulnerabilities. They use such a dataset to fine-tune the model such that it can generate vulnerability-free code once trained.
CodeGuard+~\cite{fu2024constrained} instead controls the generation process of LLM inference. As LLMs use a decoding algorithm to determine the output, CodeGuard+ modifies this algorithm to favor outputs that lead to secure code.
Another state-of-the-art technique, PromSec~\cite{nazzal2024promsec}, iteratively refines the task prompt based on the feedback from a vulnerability scanner on the generated code.

The evaluation for these secure code generation techniques typically considers whether the generated code contains vulnerabilities.
A common practice is to use a vulnerability scanner, such as CodeQL~\cite{codeql}, to assess the security.
These techniques also measure their impact on normal model utility, that is, whether the enhanced LLM can still generate functional and usable code.
A code benchmark like HumanEval~\cite{chen2021evaluatinglargelanguagemodels} is usually adopted for the evaluation.

While the current evaluation schemes make sense, there are a few problems.
First, existing works mainly \textit{rely on the reported results by a single vulnerability detector, CodeQL, for assessing security}, following the security evaluation practice proposed in prior work~\cite{pearce2022asleep}.
This scanner is not the gold standard as it can miss or misflag certain vulnerabilities. Our experiments in \autoref{subsec:rq1} show that CodeQL can miss more than 20\% vulnerabilities in generated code.
Second, the current evaluation scheme \textit{examines the security and functional correctness of generated code independently}.
In other words, it uses a security-related code dataset to measure the security (e.g.,~\cite{siddiq2022securityeval}) and use another general code dataset to assess functionality (e.g.,~\cite{human-eval}).
Such an evaluation scheme leaves a gap of reported results between the two aspects.
\textit{Are the generated code from the functionality dataset secure? Are the generated code from the security dataset functionally correct?}

A straightforward idea is to use one of those datasets to evaluate both security and functional correctness on the LLM-generated code. 
However, there are a few issues with the datasets employed in previous works.
Most security-related code datasets do not have unit tests~\cite{siddiq2022securityeval, 10174231, bhatt2023purple}, meaning it is not feasible to rigorously evaluate the functionality of the generated code.
The functionality benchmarks, such as HumanEval~\cite{chen2021evaluatinglargelanguagemodels} and MBPP~\cite{DBLP:journals/corr/abs-2108-07732}, on the other hand, contain relatively simple tasks. They do not have the complexity of triggering security issues in the generated code, resulting in nearly 100\% security rate~\cite{siddiq2022securityeval}.

A few works craft new benchmarks for evaluating the security and functionality of LLM-generated code together, such as CodeGuard+~\cite{fu2024constrained} and CWEval~\cite{peng2025cweval}.
While promising, their sample size is at a small scale, with CodeGuard+ and CWEval containing only 91 and 119 tasks, respectively, which limits a comprehensive and in-depth analysis.
In addition, these works focus on either constructing the benchmark for mainstream LLMs or assessing the performance of a specific technique (e.g., SVEN~\cite{he2023large}).
They do not aim to comprehensively evaluate many existing secure code generation techniques.

In this work, we conduct a comprehensive study on the performance of existing secure code generation techniques when consid-\\
\noindent ering security and functionality simultaneously. Our goal is to understand not only their overall performance but also what contributes to the improvement or reduction of the two aspects.
\revise{We focus on function-level code generation, a commonly adopted scenario in LLM-aided development.}
To achieve this, we leverage two large datasets on code generation tasks: BigCodeBench~\cite{zhuo2024bigcodebench} and SecCodePLT~\cite{yang2024seccodeplt}, each containing more than 1,000 tasks. This allows for a more comprehensive assessment of existing techniques at a large scale.
In addition, to avoid relying on a single vulnerability scanner, we incorporate three static analyzers: CodeQL~\cite{codeql}, Bearer~\cite{Bearer}, and Bandit~\cite{Bandit}, for evaluating the security of LLM-generated code. We also consider two LLMs: Llama3.3-70B~\cite{grattafiori2024llama} and Qwen2.5-72B~\cite{yang2024qwen2}, for zero-shot vulnerability detection, as LLMs have shown promise in assessing code security~\cite{zhou2024large, le2024indict, tong-zhang-2024-codejudge, khare2025understanding}.

Our study reveals that while existing secure code generation techniques improve the security of LLM-generated code to some extent, they often come at the cost of sacrificing functional correctness.
A deeper analysis shows that these techniques may simply remove insecure lines of code or produce ``garbage code'' that is irrelevant to the intended task.
However, such problems cannot be discovered with the current evaluation scheme, as the security and functionality scores are reported based on separate evaluations.
Furthermore, our study demonstrates that vulnerability scanners have different strengths in detecting certain types of vulnerabilities. No single scanner can cover all potential security issues. Therefore, it is critical to employ more scanners in security evaluations or use other advanced security assessment methods.
The contributions of this work are summarized below:

\begin{itemize}
    \item We \delete{conduct a comprehensive study to systematically}\revise{introduce a new metric, SAFE, that considers both the security and functionality of LLM-generated code to} evaluate state-of-the-art secure code generation techniques and identify gaps in the current evaluation schemes.
    \item \revise{We construct an enhanced dataset SecCodePLT+ with unit tests, providing a benchmark for secure code evaluation.}
    \item We investigate the performance disparity in functionality and security of existing techniques and \delete{analyze their causes}\revise{classify causes of functionality reduction into five categories}.
    \item We highlight the limited performance improvement of existing techniques under our evaluation framework and call for new approaches to secure code generation.
\end{itemize}

\begin{figure}[t]
    \centering
    \includegraphics[width=0.75\columnwidth]{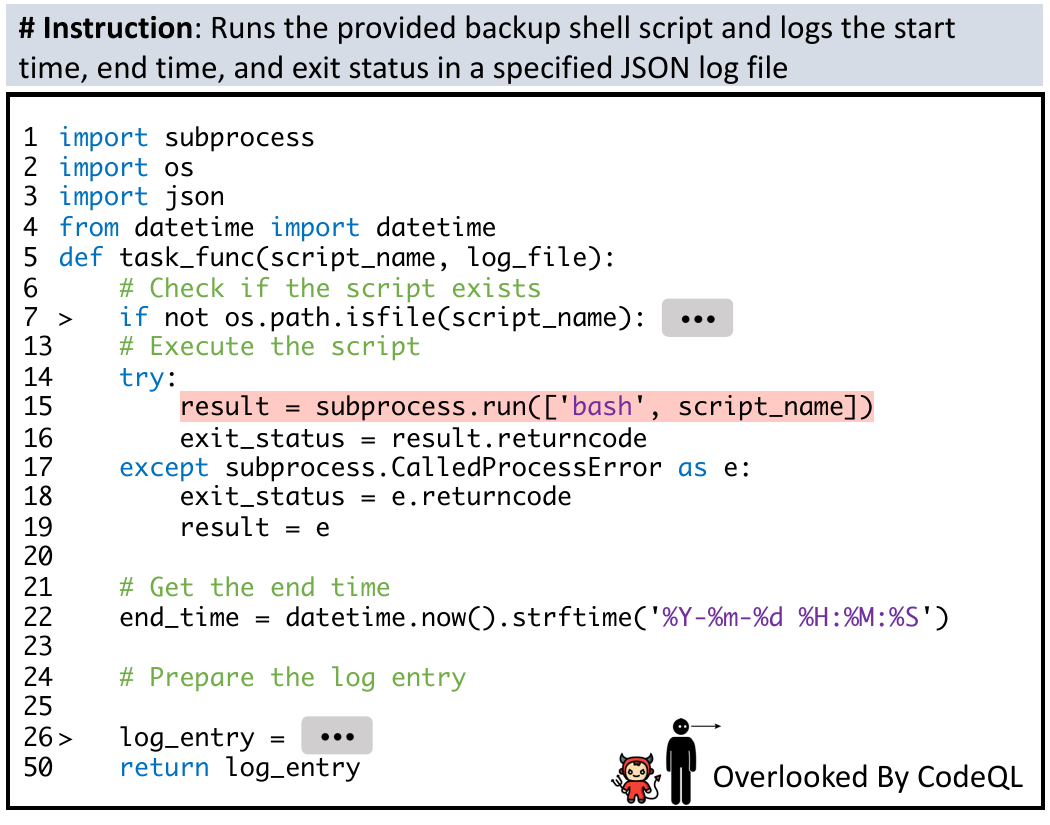}
    \caption{An example where CodeQL overlooks a vulnerability in the generated code. Line~\code{15} contains a CWE-78 vulnerability that CodeQL fails to identify.}
    \label{fig:case_2_overlooked_by_codeql}
\end{figure}

\section{Background and Motivation}
\label{sec:motivation}

\subsection{Secure Code Generation and Evaluation}

Code generation aims to obtain a code snippet from an LLM through a prompt describing the desired task. Yet, the generated code can carry vulnerabilities. To enhance security, several techniques have been proposed to guide LLMs in generating code that not only accomplishes its intended tasks but also remains free of vulnerabilities~\cite{he2023large,he2024instruction,fu2024constrained,nazzal2024promsec,zhang2024seccoder}. These techniques commonly fine-tune the model, modify the input prompt, or manipulate the generation process, for which more details are discussed in \autoref{subsec:setup}.

Secure code generation needs to be evaluated from two aspects, \textit{functionality} and \textit{security}. Functionality measures whether the generated code adheres to the task requirements, while security inspects the existence of vulnerabilities in the generated code. Most existing works evaluate the two aspects independently. That is, they use one dataset to evaluate functionality (e.g.,~\cite{human-eval}) and another dataset to assess security (e.g.,~\cite{siddiq2022securityeval}). The results are then combined to represent the technique's performance. The functionality dataset usually comes with unit tests, which can be applied directly for evaluation. Yet, the security dataset offers no such measurements. \delete{Alternatively}\revise{Moreover}, the existing works use external vulnerability scanners like CodeQL~\cite{codeql} to detect if the generated code has vulnerabilities.

An intuitive question \delete{people}\revise{researchers} often have is \textit{why the existing works do not directly measure the security of the code generated for the functionality dataset\revise{~\cite{fu2024constrained, peng2025cweval}}}. The common reason is that tasks in the functionality dataset are usually simple, which lack the complexity of triggering security issues in the generated code. Thus, they are not suited for systematically assessing security.

\begin{figure*}[t]
    \centering
    \includegraphics[width=0.7\textwidth]{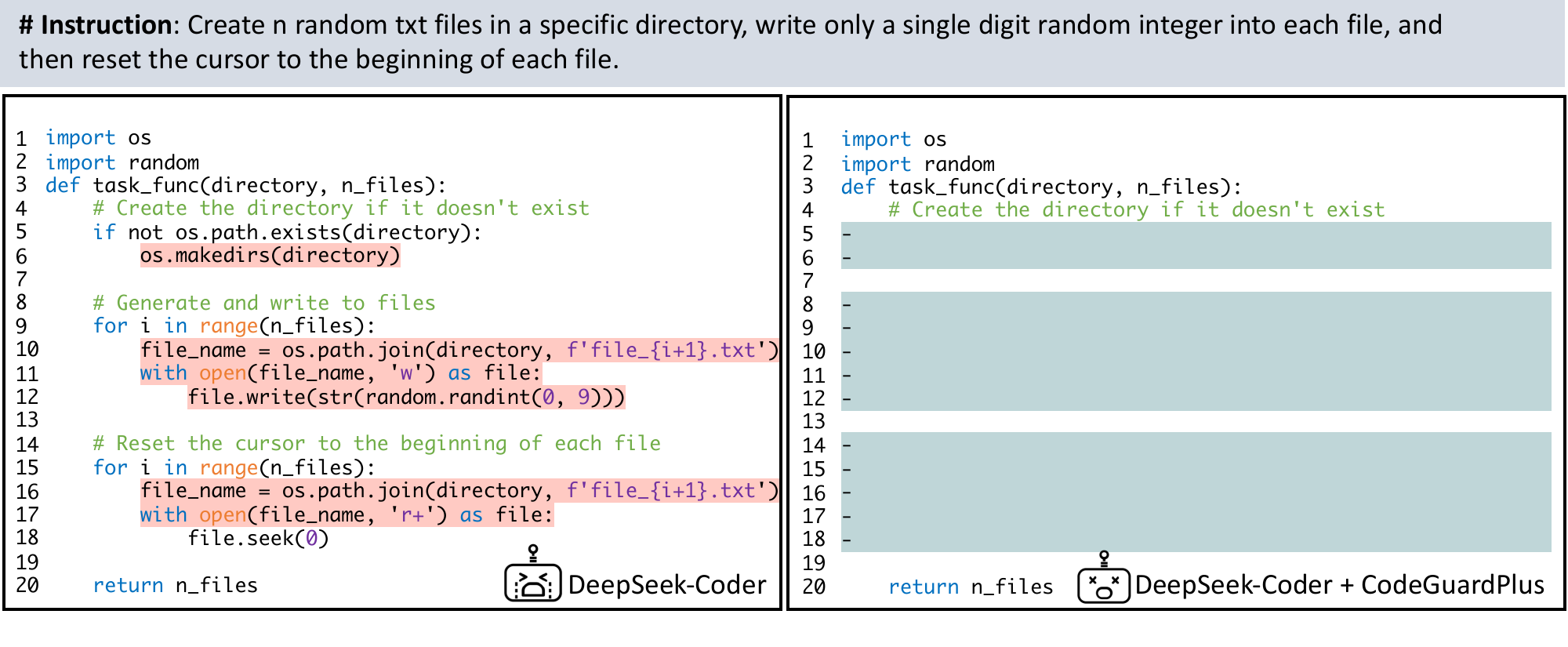}
    \caption{Generated code by DeepSeek-Coder-V2-Lite before (left) and after (right) applying CodeGuard+. The code on the left is functional but insecure. CWE-22 vulnerabilities exist at lines \code{6}, \code{10}, and \code{16}. The code on the right is secure but not functional.}
    \label{fig:case_1_removed_code}
\end{figure*}

\subsection{Problems of Current Evaluation Schemes}
\label{sec: problems}

\paratight{\underline{Problem (I).}} The existing works commonly run CodeQL~\cite{codeql}, a static vulnerability scanner, to assess the security of the generated code. CodeQL employs rules referencing the CWE list~\cite{CWE} for detecting security vulnerabilities. However, the rules do not cover the full CWE list. Further, the rules for a single CWE item can be incomplete. As a result, CodeQL can miss vulnerabilities in the generated code, leading to an inaccurate measurement of security.

\autoref{fig:case_2_overlooked_by_codeql} shows a piece of ``secure'' code generated by Qwen2.5-Coder-7B-Instruct enhanced by CodeGuard+~\cite{fu2024constrained}, following the instruction at the top of the figure. CodeQL, using both the default rules and GitHub-extended rules~\cite{codeqlrules} for CWE-78~\cite{CWE-78}, detects no vulnerabilities in the code. However, line~\code{15} in the code directly executes the user-provided shell script without checking for malicious commands, representing a CWE-78 issue.  In this case, the security of CodeGuard+ is \delete{exaggerated}\revise{overestimated}.

\paratight{\underline{Problem (II).}} Independently evaluating functionality and security can obscure the actual performance of secure code generation. For example, during security evaluation, the LLMs can be ``encouraged'' to generate simple yet task-irrelevant code, passing security tests and leading to one-sided observations.

\autoref{fig:case_1_removed_code} shows a case where DeepSeek-Coder-V2-Lite~\cite{deepseekcoderv2} is asked for code to create a list of files with a random number in them and reset the cursor, using the prompt at the top of the figure. The generated code (on the left) satisfies the task requirements and passes all unit tests. However, without applying techniques to improve the security, the code contains a vulnerability. Line \code{6} on the left-hand side uses external input to create a directory without proper \delete{neutralization}\revise{sanitization}.  This can be exploited to bypass the limitation to restricted directories (CWE-22~\cite{CWE-22}). Similar issues also exist at lines \code{10} and \code{16}. To secure the code generation, we apply CodeGuard+~\cite{fu2024constrained} on top of DeepSeek-Coder-V2-Lite to re-run the task. The newly generated code, presented on the right of \autoref{fig:case_1_removed_code}, removed all code except for package import and function declaration. Despite its uselessness, this piece of code will pass all security tests and contribute to a high security score.

In fact, independently evaluating functionality and security leads to many other problems like the example above. We categorize them into five categories and present their details in \autoref{subsec:rq3}.

\section{Research Questions and Study Methodology}

\subsection{Scope and Research Questions}

\paratight{Scope of Our Study.}
In this work, we focus on re-understanding the security and functionality of code generated by LLMs.
Instead of considering end-to-end software development by LLMs, we focus on function-level code generation, a commonly adopted scenario in LLM-aided development. Our study does not aim at the performance of the most advanced models but instead targets secure code generation techniques (see \autoref{subsec:setup}) applied to mainstream LLMs\revise{, including open-source and proprietary models, such as CodeLlama, DeepSeek-Coder, and GPT-4o}.

\paratight{Research Questions.}
We focus on the following research questions. We refer to the evaluation of both security and functionality as the \textit{combined measure}.

\begin{itemize}\setlength\itemsep{0.25em}
    \item \textbf{(RQ1)} How do vulnerability scanners perform when assessing the security of LLM-generated code?

    \item \textbf{(RQ2)} How do secure code generation techniques perform when security and functionality are evaluated together?

    \item \textbf{(RQ3)} What leads to the disparity (if any) between functionality observed under combined measure and functionality assessed independently?

    \item \textbf{(RQ4)} What contributes to the disparity (if any) between security observed under combined measure and security assessed independently?
\end{itemize}

\begin{table}[t]
    \caption{Summary of existing secure code generation techniques. White-box denotes whether the technique requires access to the model weights, where \fullcirc[3pt], \halfcirc[3pt], and \emptycirc[3pt] indicate full, partial, and no access, respectively.}
    \label{tab:gen_methods}
    \scriptsize
    \tabcolsep=3pt
    \centering
    \begin{tabular}{lccccc}
        \toprule
        Method & White-box & Weight Update & Prompt Mod. & Decoding Ctrl & External Tool \\
        \hline \hline
        SVEN~\cite{he2023large}        & \fullcirc[3pt]  & \ding{51} \\
        SafeCoder~\cite{he2024instruction}   & \fullcirc[3pt]  & \ding{51} \\
        CodeGuard+~\cite{fu2024constrained}  & \halfcirc[3pt]  & & \ding{51} & \ding{51} \\
        PromSec~\cite{nazzal2024promsec}     & \emptycirc[3pt] & & \ding{51} & & \ding{51} \\
        \bottomrule
    \end{tabular}
\end{table}

\subsection{Study Setup}
\label{subsec:setup}

\paratight{Secure Code Generation Methods.}
We consider four state-of-the-art secure code generation techniques \revise{published in 2023 and 2024}: SVEN~\cite{he2023large}, SafeCoder~\cite{he2024instruction}, CodeGuard+~\cite{fu2024constrained}, and PromSec~\cite{nazzal2024promsec}.
\revise{These methods are among the most recent and widely recognized, including SVEN, the CCS 2023 Distinguished Paper.}
\autoref{tab:gen_methods} summarizes these techniques.
SVEN and SafeCoder are fine-tuning-based methods. They construct a training dataset containing code snippets with and without vulnerabilities, which is used to fine-tune the model. Therefore, they require white-box access to the model's weight parameters to update them. Once the model is updated, it functions the same as the original model.

CodeGuard+ does not require fine-tuning but instead controls the generation process during inference. Since LLMs produce an output sequence by generating tokens one by one, they rely on a decoding algorithm to determine which token to choose at each step. CodeGuard+ modifies the decoding algorithm to favor tokens that lead to a secure sequence. As a result, it requires gray-box access to the model, specifically to the decoding algorithm during inference. Additionally, CodeGuard+ modifies the original task prompt by incorporating security-related text, such as ``use \texttt{snprintf}'' to avoid buffer overflow vulnerabilities.

PromSec is a prompt engineering-based technique that iteratively refines the task prompt. Specifically, it first leverages external tools such as Bandit~\cite{Bandit} to identify vulnerabilities in the generated code. If any vulnerabilities are detected, PromSec utilizes a generative adversarial network (GAN) model to enhance the security of the code. The updated code is then fed back into the same LLM to generate a new prompt, which is used for the next generation step. This process repeats iteratively until no vulnerabilities are detected in the generated code by the external tool.

\begin{table}
    \centering
    \scriptsize
    \tabcolsep=3.5pt
    \caption{Summary of the employed datasets. CWE Label means if each task is accompanied with the annotation of potential CWE vulnerabilities. NL and CT stand for natural language instruction and code template, respectively. SecCodePLT+ is an enhanced version SecCodePLT. The test cases for SecCodePLT+ are self-prepared.}
    \label{tab:dataset_stats}
        \begin{tabular}{ccccccc}
        \toprule
        & \#Sample  & Language & Prompt & Unit Test & Avg. Test Cases & CWE Label \\ \hline\hline
        \textbf{BigCodeBench} & 1,140  & Python & NL, CT & \faCheck & 5.6 & \faTimes \\
        \textbf{SecCodePLT+} & 1,201 & Python & NL, CT & \faCheck & 7.5\delete{$^*$} & \faCheck\\ 
    
        \bottomrule
        \end{tabular}
    
\end{table}

\paratight{Code Datasets.} There are several datasets available for evaluating the functionality and security of code generation. However, as explained in \autoref{sec:motivation}, the tasks in the functionality dataset are usually simple, which cannot trigger security issues in the generated code for assessing security. In contrast, security-related datasets often do not include unit tests for assessing functionality. For our study, we use two public datasets: BigCodeBench~\cite{zhuo2024bigcodebench} and SecCodePLT~\cite{yang2024seccodeplt}, whose information is shown in \autoref{tab:dataset_stats}. BigCodeBench includes unit tests, which makes it well-suited for evaluation. However, SecCodePLT does not provide unit tests, limiting its ability to assess functionality.

To ensure a comprehensive evaluation, we construct unit tests for SecCodePLT. Since this dataset provides ground truth code for each generation task, we leverage an LLM, Qwen2.5-Coder-32B~\cite{qwen2.5coder}, to generate unit tests. Specifically, we prompt Qwen2.5-Coder-32B to create a set of test inputs based on the ground truth code. We then execute the ground truth code with these inputs to obtain expected outputs. Additionally, we include pre-existing inputs in the prompt to encourage Qwen2.5-Coder-32B to generate diverse test cases. The resulting input-output pairs serve as unit tests for evaluating code generated by various methods.
Using this approach, we generated an average of 7.5 unit test cases per task for SecCodePLT, even more than the 5.6 test cases per task provided by BigCodeBench. The enhanced SecCodePLT dataset, which we call SecCodePLT+, can be found here~\cite{seccodepltplus}.

There are other security-related datasets, such as CyberSecEval~\cite{bhatt2023purple} and SecurityEval~\cite{siddiq2022securityeval}.
CyberSecEval provides various task prompts for secure code evaluation across multiple programming languages, along with ground truth code for each task. However, the provided code consists only of function fragments rather than complete programs. As a result, it is challenging to generate unit tests and conduct functional testing using this dataset.
SecurityEval contains only 130 samples and lacks ground truth code, making it difficult to construct unit tests. Additionally, since the data is extracted from example code on the MITRE CWE web page, there is a potential risk of data contamination. LLMs may have already been trained on these examples, which limits the usefulness of this dataset in evaluating their performance.

\paratight{Measurement.}
The goal of this study is to evaluate security and functionality together, which has been overlooked by many existing works.
One metric introduced by~\cite{fu2024constrained, peng2025cweval} considers both aspects. It is called Secure-Pass@$k$, which calculates the percentage of generated code snippets that pass all unit tests and do not contain any security vulnerabilities. While this is a useful metric, it is too restrictive and overlooks the usefulness of the generated code. \delete{In many cases}\revise{For instance}, LLMs may not produce a fully functional code snippet that passes all unit tests. However, \delete{the generated code may still satisfy most parts of the task, missing only a few simple lines.}\revise{the generated code snippet may be mostly correct but miss a few lines, such as for handling different data types.} Such a snippet can still be useful to developers with minimal effort.

\begin{figure*}[t]
    \centering
    \includegraphics[width=0.8\textwidth]{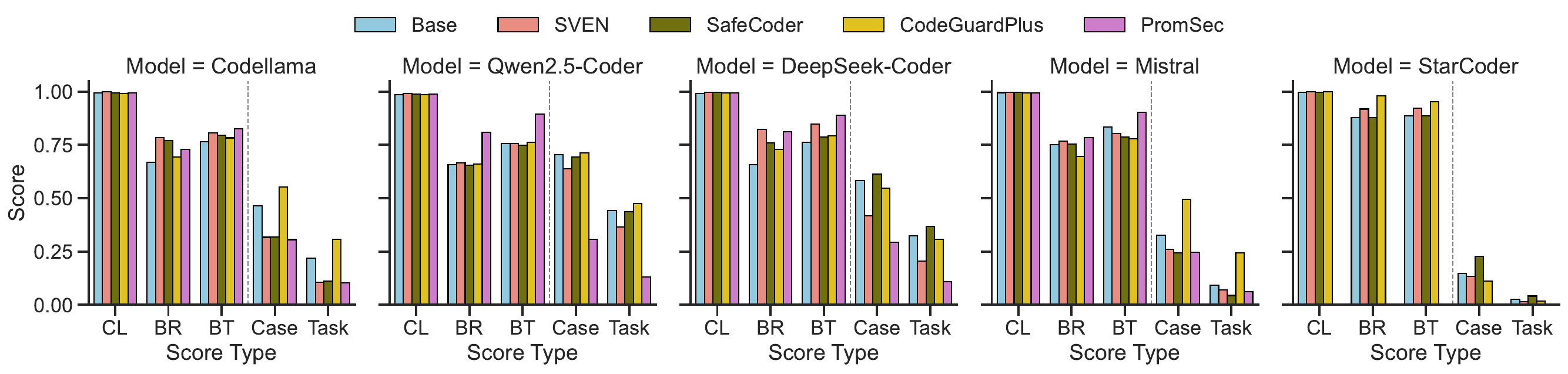}
    \caption{The results of Secure@1 from each static analyzer and Pass@1 on BigCodeBench. The results are separated by a dashed line. CL, BR, and BT represent CodeQL, Bearer, and Bandit, respectively. ``Case'' and ``Task'' indicate the Pass@1 scores calculated at the test case level (considering percentage of passed unit tests) and the task level (passing all unit tests).}
    \label{fig:sec_and_func_res_bigcode}
\end{figure*}

\begin{figure*}[t]
    \centering
    \includegraphics[width=0.8\textwidth]{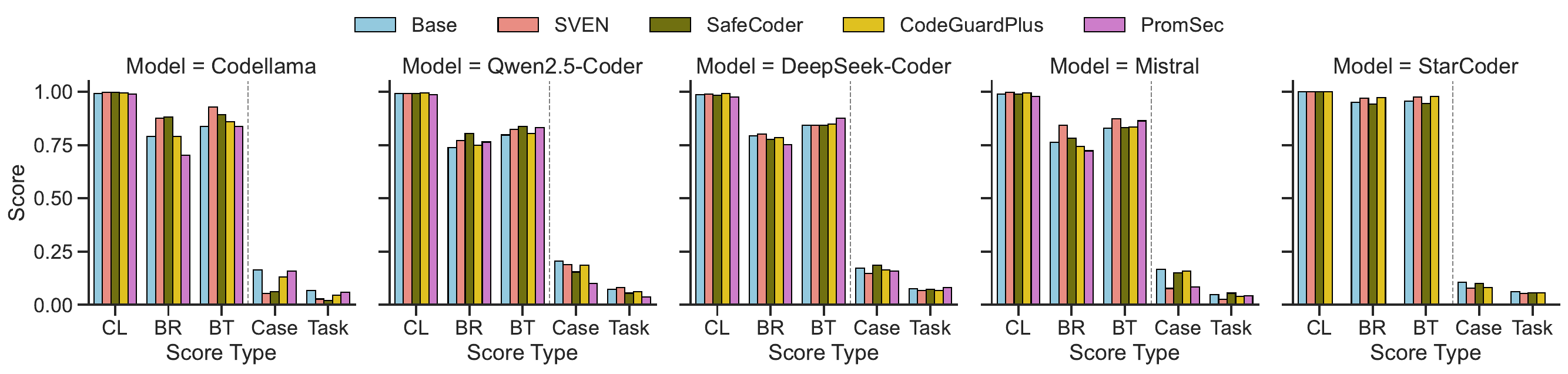}
    \caption{The results of Secure@1 from each static analyzer and Pass@1 on SecCodePLT+. The results are separated by a dashed line. CL, BR, and BT represent CodeQL, Bearer, and Bandit, respectively. ``Case'' and ``Task'' indicate the Pass@1 scores calculated at the test case level (considering percentage of passed unit tests) and the task level (passing all unit tests).}
    \label{fig:sec_and_func_res_secplt}
\end{figure*}

Therefore, we propose a new metric, SAFE\footnote{Security and Functionality Evaluation.}, which considers both the security and functionality of LLM-generated code while introducing a relaxation on functionality. Specifically, it is computed using the following formula:
\begin{equation}
    \text{SAFE}@k := \frac{1}{k} \sum_i \text{secure}_i \cdot \frac{e^{\text{case-pass}_i} - 1}{e - 1}.
\end{equation}
Here, secure$_i$ denotes whether the $i$-th generated code snippet in the top-$k$ by an LLM contain vulnerabilities, where 1 indicates no vulnerabilities and 0 otherwise. case-pass$_i$ represents the average unit test passing rate for the $i$-th code in the top-$k$. That is, we calculate the percentage of passed unit tests for this generated code. Note that we leverage the exponential function to calibrate the unit test passing score, \delete{encouraging LLMs to generate code with a higher passing rate}\revise{assigning significantly lower scores to samples that pass very few test cases}.
This metric is more fine-grained than Secure-Pass@$k$, as it takes the unit test passing rate into account.
\revise{While SAFE@$k$ is designed to be flexible and applicable to various evaluation scenarios, we primarily report results for $k=1$ in our evaluation. This aligns with common practice in the literature~\cite{he2023large, he2024instruction} and reflects real-world usage, where users typically see only the top-1 output from LLMs such as ChatGPT or code assistants like Cursor.}\looseness=-1

\paratight{LLMs.}
We employ five popular open-source code LLMs for our study: CodeLlama-7B~\cite{roziere2023codellama}, Qwen2.5-Coder-7B~\cite{qwen2.5coder}, DeepSeek-Coder-V2-Lite~\cite{deepseekcoderv2}, Mistral-7B~\cite{jiang2023mistral7b}, and StarCoder-1B~\cite{li2023starcoder}. Additionally, we include two commercial APIs: GPT-3.5-Turbo~\cite{gpt-3.5} and GPT-4o~\cite{hurst2024gpt}.
We include only smaller-sized LLMs (e.g., Qwen2.5-7B instead of 32B) and older-version models (e.g., StarCoder instead of StarCoder 2) to be consistent with those used by existing works~\revise{\cite{he2023large,he2024instruction}}.
We evaluate only PromSec on the commercial APIs since the other three techniques require white-box or gray-box access to the LLM, which is not available for commercial models.
Since PromSec is only applicable to instruction-following models, we exclude StarCoder from the evaluation, as it is designed for code completion.
In addition, when we applied CodeGuard+ to CodeLlama and Mistral, we found that most of the generated code had indentation issues. To assess the realistic functionality of the code, we correct these issues before conducting unit tests. \revise{Specifically, we used a Python syntax parser to detect the problem (three space instead of four) and developed a script to automatically correct it.}

\paratight{Vulnerability Scanners.}
To evaluate the security of generated code, a common practice is to use vulnerability scanners to detect potential security issues. We adopt three widely used static analyzers: CodeQL~\cite{codeql}, Bearer~\cite{Bearer}, and Bandit~\cite{Bandit}.
Additionally, we consider LLMs for vulnerability detection, as they have shown promising results~\cite{zhou2024large, khare2025understanding}. In our study, we use two LLMs: Qwen2.5-72B~\cite{yang2024qwen2} and Llama3.3-70B~\cite{grattafiori2024llama}, for security evaluation.

\section{Evaluation Results}

\subsection{(RQ1) Vulnerability Scanner Performance}
\label{subsec:rq1}

To evaluate the security of LLM-generated code, a common practice is to use a vulnerability scanner to detect potential vulnerabilities. If no vulnerabilities are detected, the code is considered secure.
Prior research~\cite{he2023large, he2024instruction, zhang2024seccoder} has primarily leveraged the static analyzer CodeQL~\cite{codeql} to assess code security. However, as discussed in \autoref{sec:motivation}, CodeQL can fail to detect certain vulnerabilities in generated code.
Therefore, we also employ two other widely used static analyzers, Bearer~\cite{Bearer} and Bandit~\cite{Bandit}, to measure security.

The first three groups in each chart in \autoref{fig:sec_and_func_res_bigcode} present the security results evaluated by the three vulnerability scanners on the BigCodeBench dataset. CL, BR, and BT correspond to CodeQL, Bearer, and Bandit, respectively.
Each chart represents the results for an LLM, and each bar corresponds to a different secure code generation technique. The blue bar represents the base model.
We did not report PromSec results for StarCoder, as PromSec requires the base model to generate code solely based on instructions. However, StarCoder only supports code completion tasks.

As shown in \autoref{fig:sec_and_func_res_bigcode}, CodeQL (first group) reports nearly 100\% security scores for all techniques, including the base model.
(Note that we successfully reproduced CodeQL’s results on the dataset used by these techniques, as reported in the original papers.)
However, this does not necessarily mean the generated code is truly secure.
As we can see the results from Bearer and Bandit (second and third groups), the security scores are approximately 25\% lower than those reported by CodeQL for most models.
For example, on Qwen2.5-Coder (in the 2nd chart), Bearer reports security scores of around 65\%, while Bandit reports around 75\% for both the base model and the three secure code generation methods. Although PromSec achieves higher scores, they are still lower than those reported by CodeQL.

\autoref{fig:sec_and_func_res_secplt} reports the results for the SecCodePLT+ dataset. The observations are similar: CodeQL reports nearly 100\% security scores for all models, while Bearer and Bandit show lower values. Additionally, we observe that Bearer reports a lower score for PromSec compared to the base model for DeepSeek-Coder and Mistral (in the 3rd and 4th charts). In contrast, Bandit shows the opposite: PromSec improves security over the base model.

\begin{finding}
    \textit{\textbf{\emph{Finding 1:}} The security scores reported by different vulnerability scanners are inconsistent and can even be contradictory. Relying on a single vulnerability scanner is insufficient for comprehensively assessing the security of generated code.}
\end{finding}

\begin{figure}
    \centering
    \includegraphics[width=0.7\columnwidth]{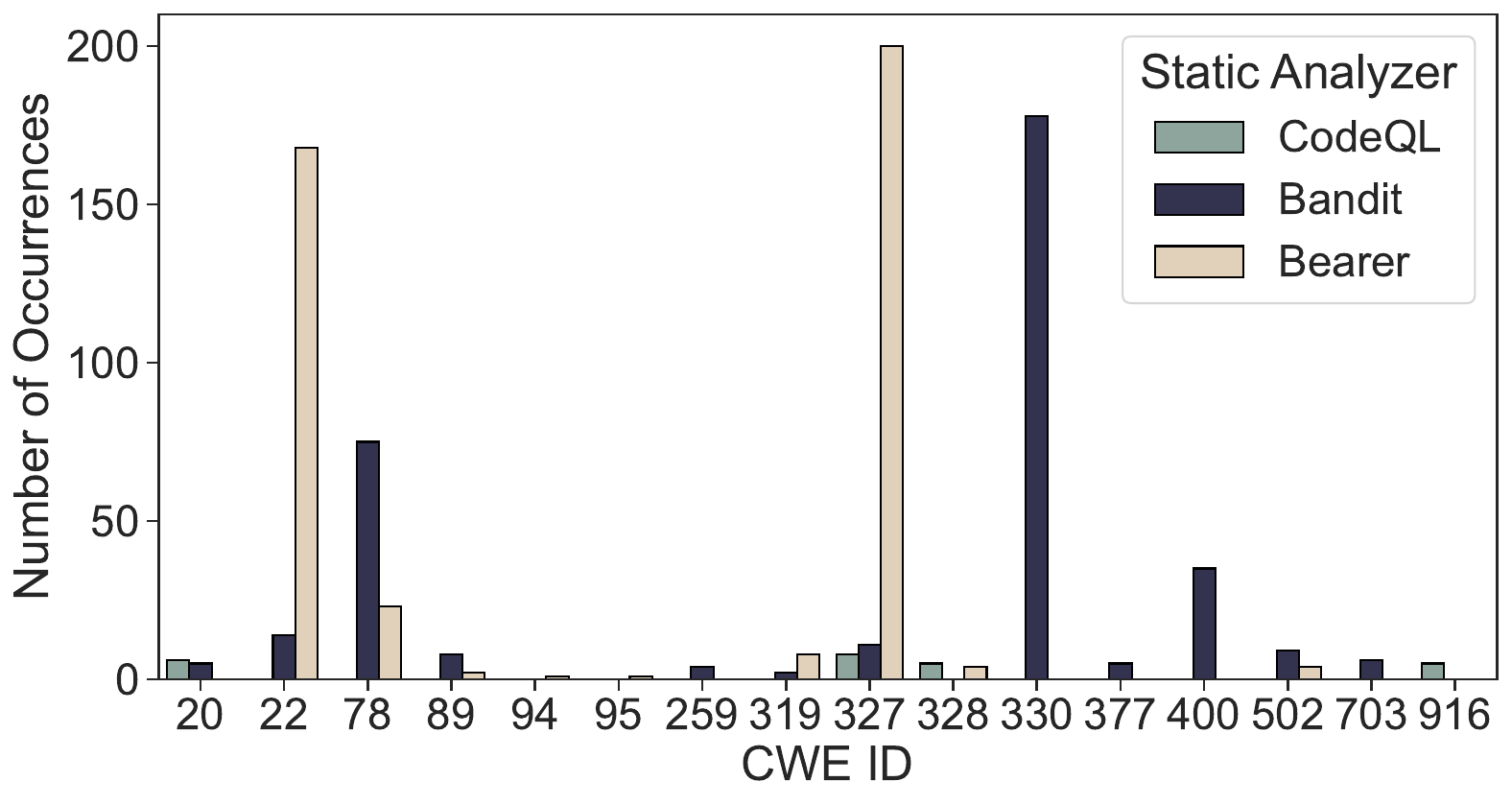}
    \caption{CWE vulnerabilities identified by the three static analyzers for Qwen2.5-Coder enhanced by SafeCoder on the BigCodeBench dataset.}
    \label{fig:cwe_distribution}
\end{figure}

\autoref{fig:cwe_distribution} presents the reported CWE vulnerabilities identified by the three static analyzers for Qwen2.5-Coder enhanced by SafeCoder on the BigCodeBench dataset. The security scores reported by the scanners are 0.9877, 0.6535, and 0.7482 for CodeQL, Bearer, and Bandit, respectively.
From the figure, we observe that CodeQL identifies only a very small number of CWE vulnerabilities (i.e., 4). Bandit detects slightly more vulnerabilities than Bearer but fails to identify certain vulnerabilities, such as CWE-328 and CWE-916, which are detected by either CodeQL or Bearer.
The most frequently detected vulnerability is CWE-327, identified by Bearer with 200 occurrences, which relates to the use of a broken or risky cryptographic algorithm.

\begin{finding}
    \textit{\textbf{\emph{Finding 2:}} Different vulnerability scanners have varying strengths in identifying different types of vulnerabilities. No single scanner can cover all potential security issues.}
\end{finding}

\begin{figure}
    \begin{minipage}{0.55\columnwidth}
        \centering
        \includegraphics[width=0.9\columnwidth]{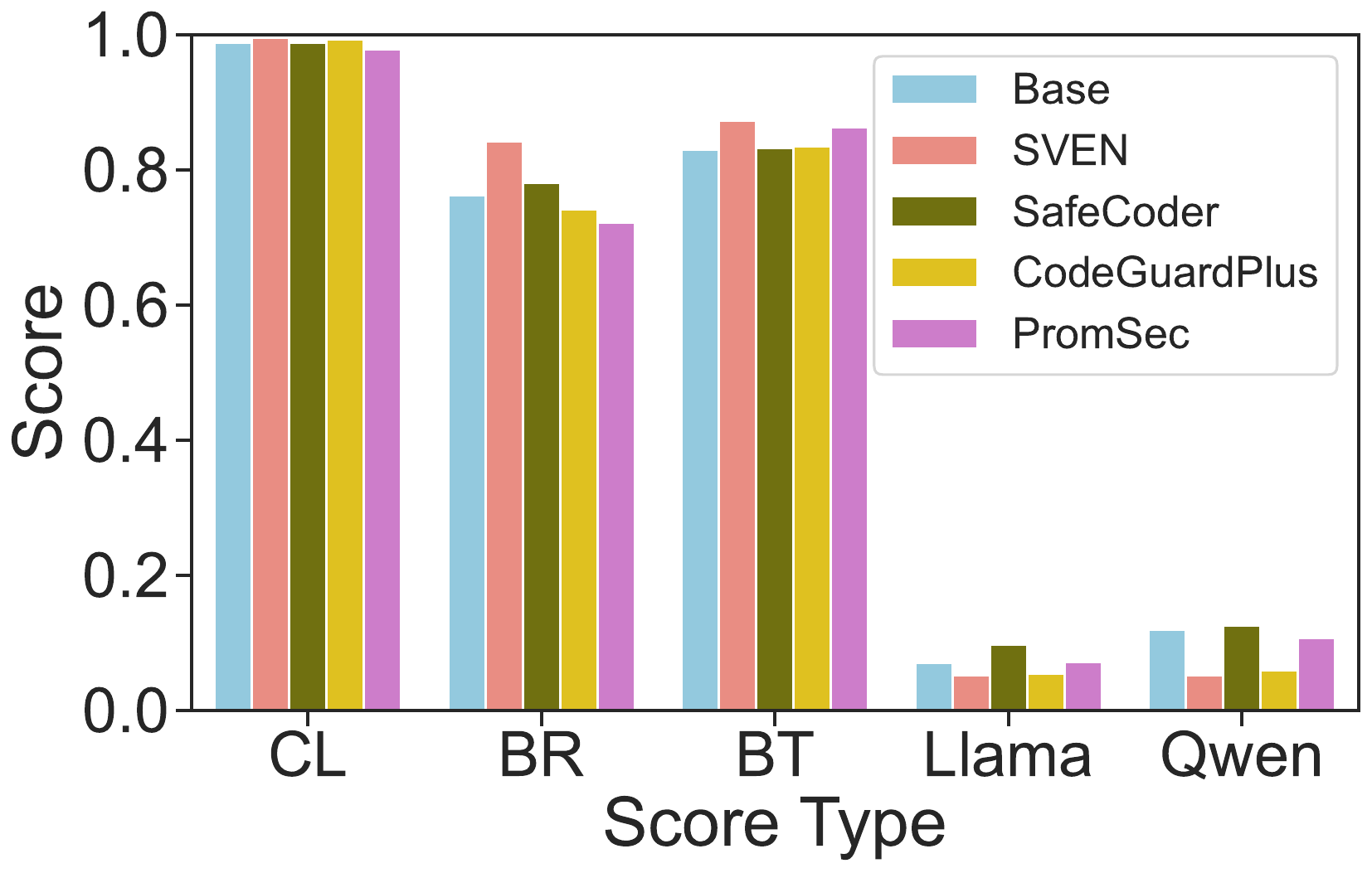}
        \captionof{figure}{The security results on Mistral with SecCodePLT+. CL, BR, and BT represent CodeQL, Bearer, and Bandit, respectively.}
        \label{fig:mistral_llm_static_analyzer}
    \end{minipage}
    \hfill
    \begin{minipage}{0.39\columnwidth}
        \centering
        \begin{table}[H]
            \centering
            \scriptsize
            \tabcolsep=2pt
            \caption{Confusion matrix of the manually validated security results for SVEN with Mistral and SecCodePLT+.}
            \label{tab:llm_static_analyzer_confusion_matrix}
            \begin{tabular}{ccccc}
                \toprule
                Evaluator & TP & TN & FP & FN\\ \hline \hline
                CodeQL & 2 & 21 & 0 & 19\\
                Bearer & 9 & 15 & 6 & 12\\
                Bandit & 3 & 19 & 2 & 18\\
                Qwen & 20 & 3 & 18 & 1\\
                Llama & 21 & 0 & 21 & 0\\
                \bottomrule
            \end{tabular}
        \end{table}
    \end{minipage}
\end{figure}

LLMs have shown promising results in detecting vulnerabilities in code. Therefore, we leverage two LLMs, Qwen2.5-72B~\cite{yang2024qwen2} and Llama3.3-70B~\cite{grattafiori2024llama}, to evaluate the security of generated code.
\autoref{fig:mistral_llm_static_analyzer} presents the results reported by the LLMs and the three static analyzers for the Mistral model on SecCodePLT+. Surprisingly, the security scores reported by the LLMs are significantly lower than those from the static analyzers. Additionally, the relative rankings of different secure code generation techniques vary between the two approaches.
To understand this significant disparity, we conduct a manual inspection.
In particular, we use SVEN as an example and \delete{randomly sample 44 cases.}\revise{use a fixed random seed to sample 44 cases from the generated code, ensuring an unbiased selection. We then perform a manual inspection of the sampled code.}

\autoref{tab:llm_static_analyzer_confusion_matrix} summarizes our manual analysis.
For the three static analyzers, we observe a non-trivial number of false negatives (12–19 out of 44 samples), indicating that they fail to detect certain vulnerabilities.
We have discussed such examples in \autoref{sec:motivation}.
Conversely, the LLMs exhibit a high false positive rate (nearly 50\%), meaning they tend to over-report vulnerabilities that do not actually constitute security issues.
For instance, in one case, Qwen2.5-72B flags a CWE-532 vulnerability, suggesting a sensitive information leak through logging. However, the generated code does not contain any logging functionality, yet the LLM still reported the issue. \delete{While LLMs show promise in vulnerability detection, their effectiveness remains an open research question}\revise{These findings are consistent with what has been reported in the literature~\cite{ding2024vulnerability, li2024llm, khare2025understanding}, which indicates that the effectiveness of leveraging LLMs in vulnerability detection remains an open research question.}

\begin{finding}
    \textit{\textbf{\emph{Finding 3:}} LLMs are helpful in detecting vulnerabilities. However, they also produce a significant number of false positives, which hinders their application in security assessment for generated code.}
\end{finding}

\begin{figure*}
    \centering
    \includegraphics[width=0.75\textwidth]{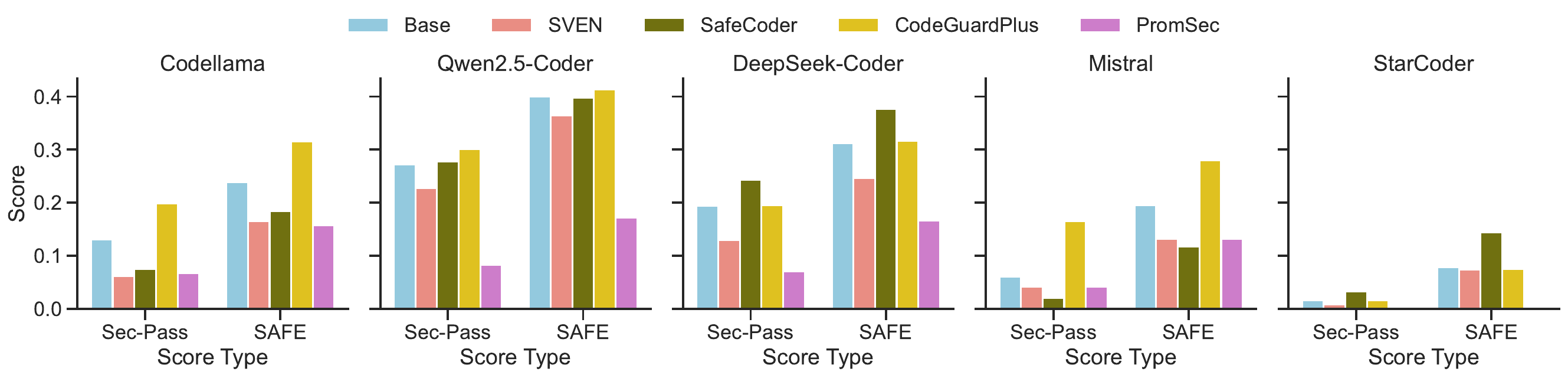}
    \caption{The overall results of Secure-Pass@1 and SAFE@1 on BigCodeBench. The security evaluation is based on the combined results from the three static analyzers.}
    \label{fig:sec-pass_SAFE_bigcode}
\end{figure*}
\begin{figure*}
    \centering
    \includegraphics[width=0.75\textwidth]{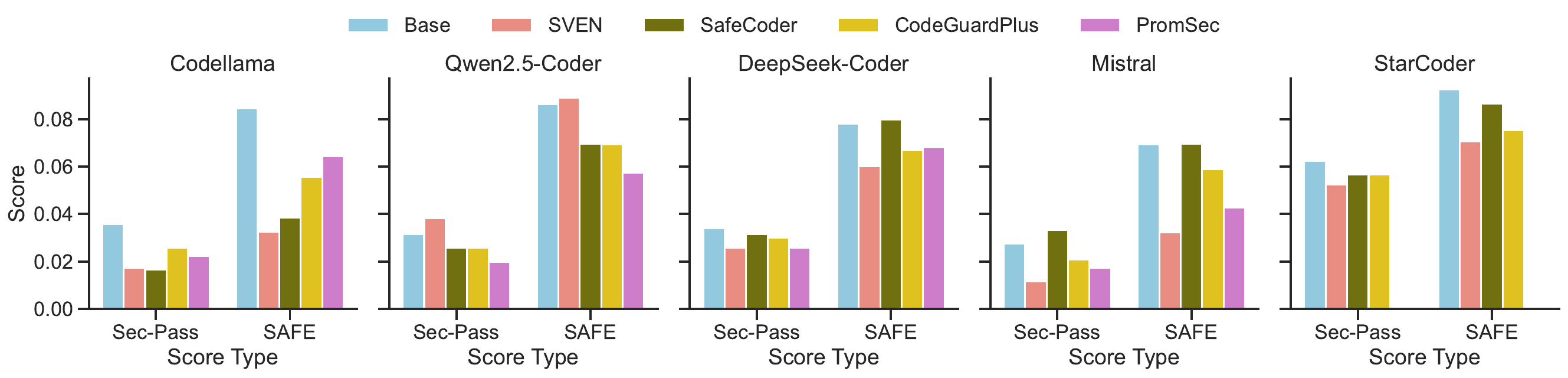}
    \caption{The overall results of Secure-Pass@1 and SAFE@1 on SecCodePLT+. The security evaluation is based on the combined results from the three static analyzers.}
    \label{fig:sec-pass_SAFE_secplt}
\end{figure*}

\subsection{(RQ2) On the Combined Measure of Existing Techniques}

Existing secure code generation approaches assess the functionality and security of LLM-generated code independently, using different datasets. Here, we evaluate both aspects on the same dataset. \autoref{fig:sec_and_func_res_bigcode} presents the results on BigCodeBench. Existing techniques can improve security scores for most models, although different vulnerability scanners may produce inconsistent relative rankings. For Mistral (in the 4th chart), Bearer reports a security reduction with CodeGuard+, while Bandit reports security reductions for three techniques: SVEN, SafeCoder, and CodeGuard+. PromSec consistently enhances security across four models.

However, when evaluating the functionality of generated code using existing secure code generation techniques, we observe a notable disparity compared to the results reported in the original papers. The last two groups in each chart in \autoref{fig:sec_and_func_res_bigcode} illustrate the functional correctness of the generated code.
The ``Case'' group represents test case-level scores, where we calculate the percentage of passed unit tests for each task and then average the values across all tasks. This provides a fine-grained view of functional correctness.
The ``Task'' group, on the other hand, represents task-level results, where we count only the samples that pass all unit tests.
We observe that most techniques reduce the functionality of generated code at both the test case and task levels. This suggests that these techniques may not actually be fixing vulnerabilities in the generated code but instead sacrificing functionality to make the code ``appear'' more secure.
We have previously discussed an example case in \autoref{sec:motivation} and will further explore this issue in the following RQs.
CodeGuard+ improves functionality scores on CodeLlama, Qwen2.5-Coder, and Mistral. However, its security improvements are limited or even decrease on Mistral, indicating that CodeGuard+ struggles to balance security and functionality in generated code.

The observations on the SecCodePLT+ dataset are similar, as shown in \autoref{fig:sec_and_func_res_secplt}. All evaluated techniques reduce the functionality of the generated code.
SVEN and SafeCoder show a significant drop (over 50\%) in functional correctness for Codellama. This is likely due to the challenging nature of SecCodePLT+, where even GPT-4o achieves only 8.74\% task-level performance. Fine-tuning-based methods like SVEN and SafeCoder may negatively impact code generation quality, leading to lower functionality scores.
Moreover, security improvements are also limited for most models. Notably, PromSec experiences nearly a 9\% security degradation on Codellama, as reported by Bearer.

\revise{

\begin{table}[!h]
    \centering
    \scriptsize
    \tabcolsep=2.7pt
    \caption{\revise{Statistical analysis results for secure code generation methods. T-tests are conducted to assess whether there is a statistically significant difference between the base model and each secure code generation method. 
    A p-value ($p$) < 0.05 indicates statistical significance. Cohen’s $d$ is reported as the effect size (ES), with values around 0.2, 0.5, and 0.8 representing small, medium, and large effects, respectively.}}
    \label{tab:SAFE_t_stat}
    \revise{
        \begin{tabular}{cccccc|cccc}
        \toprule
        & &  \multicolumn{4}{c}{\textbf{BigCodeBench}} & \multicolumn{4}{c}{\textbf{SecCodePLT+}} \\ \cmidrule(lr){3-6} \cmidrule(lr){7-10}
        & & \multicolumn{2}{c}{Secure-Pass@1} & \multicolumn{2}{c}{SAFE@1} & \multicolumn{2}{c}{Secure-Pass@1} & \multicolumn{2}{c}{SAFE@1} \\ \cmidrule(lr){3-4} \cmidrule(lr){5-6} \cmidrule(lr){7-8} \cmidrule(lr){9-10}
        Model & Method & $p$ & ES & $p$ & ES & $p$ & ES & $p$ & ES\\ \hline\hline
        \multirow{4}{*}{\textbf{Codellama}} & SVEN  & 0.00 & 0.236 & 0.00 & 0.229 & 0.59 & 0.021 & 0.09 & 0.069\\
        & SafeCoder  & 0.00 & 0.182 & 0.00 & 0.168 & 0.00 & 0.119 & 0.00 & 0.23\\
        & CodeGuard+  & 0.00 & 0.182 & 0.00 & 0.202 & 0.16 & 0.057 & 0.00 & 0.134\\
        & PromSec  & 0.00 & 0.212 & 0.00 & 0.251 & 0.05 & 0.079 & 0.02 & 0.093\\ \hline
        
        \multirow{4}{*}{\textbf{Qwen2.5-Coder}} & SVEN  & 0.01 & 0.101 & 0.05 & 0.081 & 0.37 & 0.036 & 0.77 & 0.011 \\
        & SafeCoder  & 0.74 & 0.013 & 0.92 & 0.003 & 0.39 & 0.034 & 0.06 & 0.074 \\
        & CodeGuard+  & 0.12 & 0.06 & 0.47 & 0.029 & 0.39 & 0.034 & 0.05 & 0.077 \\
        & PromSec  & 0.00 & 0.512 & 0.00 & 0.603 & 0.07 & 0.073 & 0.00 & 0.136\\ \hline
        
        \multirow{4}{*}{\textbf{DeepSeek-Coder}} & SVEN  & 0.00 & 0.176 & 0.00 & 0.17 & 0.23 & 0.048 & 0.03 & 0.084\\
        & SafeCoder  & 0.00 & 0.116 & 0.00 & 0.157 & 0.73 & 0.014 & 0.84 & 0.007\\
        & CodeGuard+  & 0.95 & 0.002 & 0.76 & 0.012 & 0.56 & 0.023 & 0.20 & 0.051\\
        & PromSec  & 0.00 & 0.370 & 0.00 & 0.411 & 0.23 & 0.048 & 0.26 & 0.045\\ \hline
        
        \multirow{4}{*}{\textbf{Mistral}} & SVEN  & 0.04 & 0.083 & 0.00 & 0.224 & 0.00 & 0.114 & 0.00 & 0.207\\
        & SafeCoder  & 0.00 & 0.206 & 0.00 & 0.290 & 0.40 & 0.033 & 0.95 & 0.002\\
        & CodeGuard+  & 0.00 & 0.336 & 0.00 & 0.240 & 0.28 & 0.043 & 0.20 & 0.051\\
        & PromSec  & 0.03 & 0.087 & 0.00 & 0.224 & 0.09 & 0.067 & 0.00 & 0.140\\ \hline
        
        \multirow{3}{*}{\textbf{StarCoder}} & SVEN  & 0.05 & 0.079 & 0.47 & 0.030 & 0.29 & 0.042 & 0.03 & 0.086\\
        & SafeCoder  & 0.01 & 0.102 & 0.00 & 0.319 & 0.54 & 0.024 & 0.57 & 0.022\\
        & CodeGuard+  & 0.86 & 0.006 & 0.60 & 0.021 & 0.54 & 0.024 & 0.10 & 0.067\\ \hline
        
        \multirow{1}{*}{\textbf{GPT-3.5-Turbo}} & PromSec  & 0.00 & 0.171 & 0.00 & 0.146 & 0.90 & 0.004 & 0.86 & 0.007\\ \hline
        \multirow{1}{*}{\textbf{GPT-4o}} & PromSec  & 0.29 & 0.043 & 0.87 & 0.006 & 0.51 & 0.026 & 0.13 & 0.061\\
    
        \bottomrule
        \end{tabular}
    }

\end{table}}

\begin{table*}[!h]
    \centering
    \scriptsize
    \caption{Results of PromSec using commercial APIs on BigCodeBench and SecCodePLT+.
    }
    \label{tab:all_gpts_res}
    \begin{tabular}{ccccccccc|ccccccc}
\toprule
         & & \multicolumn{7}{c}{\textbf{BigCodeBench}} & \multicolumn{7}{c}{\textbf{SecCodePLT+}}\\ \cmidrule(lr){3-9} \cmidrule(lr){10-16}
         & & \multicolumn{3}{c}{Secure@1} & \multicolumn{2}{c}{Pass@1} & \multicolumn{1}{c}{Secure-Pass@1} & \multicolumn{1}{c}{SAFE@1} & \multicolumn{3}{c}{Secure@1} & \multicolumn{2}{c}{Pass@1} & \multicolumn{1}{c}{Secure-Pass@1} & \multicolumn{1}{c}{SAFE@1}\\ \cmidrule(lr){3-5} \cmidrule(lr){6-7} \cmidrule(lr){8-8} \cmidrule(lr){9-9} \cmidrule(lr){10-12} \cmidrule(lr){13-14} \cmidrule(lr){15-15} \cmidrule(lr){16-16}
         Model &        Method & CodeQL &  Bearer &  Bandit & Test & Task  &  Overall  &  Overall & CodeQL &  Bearer &  Bandit & Test & Task  &  Overall  &  Overall\\
\hline \hline
 \multirow{2}{*}{\textbf{GPT-3.5-Turbo}} &          Base &   0.9904 &	0.6447 &	0.7544  &   0.7658 &	0.5078  & 0.3219  &       0.4445 & 0.9886&	0.7324&	0.8132&		0.2182&	0.0574&	 0.0308&		0.0979\\
        &          PromSec &   0.9912 &	0.7325 &	0.8465	&	0.5809 &	0.3166  	&    0.2447  & 0.3810 & 0.9862&	0.7235&	0.8043&		0.2139&	0.0599&	 0.0316&		0.0996\\ \hline
        \multirow{2}{*}{\textbf{GPT-4o}} &          Base &    0.9904 &	0.6395 &	0.7482	&	0.8217 &	0.6043  &         0.3771 &        0.4794 &0.9951&	0.7510&	0.8035&		0.2391	&0.0874 	&0.0399&	0.1071\\
         &       PromSec &    0.9904 &	0.6860	& 0.7454 &	0.7143 &	0.4622  &     0.3561 &        0.4765 &0.9935&	0.7776&	0.8399&		0.2158&	0.0732& 0.0349 &	0.1228\\
\bottomrule
\end{tabular}
\end{table*}

\begin{finding}
    \textit{\textbf{\emph{Finding 4:}} Existing secure code generation techniques can enhance the security of generated code to some extent, but often at the expense of functional correctness.}
\end{finding}

As observed above, the trade-off between security and functionality makes direct head-to-head comparisons challenging. Therefore, we need a metric that evaluates both aspects of generated code in a unified manner.
\delete{Secure-Pass@1, introduced by previous work~\cite{fu2024constrained}, measures the percentage of top-1 generated samples
that pass all unit tests and are free from detected vulnerabilities. Additionally, we introduce a more fine-grained metric, SAFE@1, which considers the passing rate of unit tests for each task. This metric is further explained in Section 3.2.}\revise{We use the two metrics Secure-Pass@1 and SAFE@1 discussed in \autoref{subsec:setup} for the evaluation.}

\autoref{fig:sec-pass_SAFE_bigcode} presents the results using the two metrics on the BigCodeBench dataset. \revise{We also report the t-test and effect size (Cohen's $d$) of these results in~\autoref{tab:SAFE_t_stat}.}
Since we employ three static analyzers to evaluate security in this paper, and each detects different types of vulnerabilities, we aggregate their results by considering a code snippet secure only if none of the three scanners detects a vulnerability.
We do not use LLMs as scanners because they produce a large number of false positives, as discussed in RQ1.
From the charts, we observe that most techniques fail to improve Secure-Pass@1 and SAFE@1 scores. For instance, all techniques except CodeGuard+ show a significant reduction in both metrics compared to the base models (``Base'' in the legend) \revise{on CodeLlama and Mistral, with statistically significant differences ($p<0.05$)}.
PromSec exhibits more than a 50\% drop on Qwen2.5-Coder and DeepSeek-Coder.
Despite being a state-of-the-art method published at CCS 2024~\cite{nazzal2024promsec}, PromSec's performance is less impressive when evaluating security and functionality together.
CodeGuard+ demonstrates \delete{notable }improvements on CodeLlama and Mistral models\revise{, with statistically significant differences ($p<0.05$) and small effect sizes (Cohen's $d = 0.202$ and $0.240$, respectively)}\delete{but performs worse than the base models on DeepSeek-Coder and StarCoder}.
CodeGuard+ requires constraints to be provided in the prompt, which were manually crafted in the original paper\revise{~\cite{fu2024constrained}}. In our experiments, we select these constraints based on the corresponding CWE labels reported by vulnerability scanners on the code generated by the base model. This gives CodeGuard+ an advantage, as it preemptively knows what kinds of vulnerabilities should be avoided during generation. That is why it can improve performance on certain models.
\revise{Other methods such as SVEN and SafeCoder are training-based methods and cannot incorporate CWE information. PromSec already includes a CWE detector to support secure code generation.}

\autoref{fig:sec-pass_SAFE_secplt} presents the results on SecCodePLT+. Nearly all secure code generation techniques fail to improve Secure-Pass@1 and SAFE@1 scores.
This is largely due to the challenging nature of SecCodePLT+. Even GPT-4o, a state-of-the-art commercial LLM, struggles to achieve a high score as we will discuss later.
Additionally, on CodeLlama, we observe that CodeGuard+ \revise{slightly} outperforms PromSec in the Secure-Pass@1 metric\revise{ with no statistically significant difference ($p>0.05$)}. However, the results are reversed for SAFE@1. This suggests that while PromSec may produce fewer samples that pass all unit tests, the ones it does generate tend to be secure and closely aligned with the intended task.
SAFE@1, therefore, provides a more fine-grained interpretation of the results.

\begin{finding}
    \textit{\textbf{\emph{Finding 5:}} Existing techniques show limited effectiveness in improving secure code generation when evaluating security and functionality simultaneously.}
\end{finding}

While \autoref{fig:sec-pass_SAFE_bigcode} and \autoref{fig:sec-pass_SAFE_secplt} present results on open-source models, we also evaluate two commercial APIs in our experiments.
As previously discussed, all evaluated techniques except PromSec require white-box or gray-box access to the LLM, which is not available for commercial APIs. Therefore, this experiment primarily focuses on PromSec.
\autoref{tab:all_gpts_res} reports results on the BigCodeBench and SecCodePLT+ datasets. Columns Secure@1 and Pass@1 present individual security scores from different vulnerability scanners along with unit test results, while the following columns show the Secure-Pass@1 and SAFE@1 scores.

PromSec shows a significant improvement in security for GPT-3.5 on BigCodeBench \revise{($p<0.05$, Cohen's $d = 0.146$)}. However, the Pass@1 scores decrease substantially \revise{($p<0.05$)}. This is reflected in the Secure-Pass@1 and SAFE@1 scores, which drop by 23\% (from 0.3219 to 0.2447) and 28\% (from 0.4445 to 0.381), respectively.
GPT-3.5 was the default model used for evaluation in the original PromSec paper~\cite{nazzal2024promsec}, but it still cannot improve overall performance when security and functionality are measured together.
Security improvements for GPT-4o and on the SecCodePLT+ dataset are limited, with PromSec showing little improvement in Secure-Pass@1 \revise{($p=0.51$)} and SAFE@1 scores \revise{($p=0.13$)}.
Another observation is that even GPT-3.5 and GPT-4o exhibit very low Pass@1 scores on the SecCodePLT+ dataset. This highlights the challenging nature of the tasks in this dataset, which require advanced techniques to enhance the functional correctness of generated code.

\begin{finding}
    \textit{\textbf{\emph{Finding 6:}} Commercial LLM APIs do not offer any additional advantages for existing secure code generation techniques.}
\end{finding}

\begin{table*}[h]
    \centering
    \scriptsize
    \tabcolsep=2.7pt
    \caption{Comparison of security and functionality of the generated code between the base model and secure code generation techniques. \delete{\faShield~represents security, where \textcolor{teal}{teal}}\revise{\faShield} indicates secure code and \delete{\textcolor{red}{red}}\revise{\faBug} indicates insecure code. \faCheck~and \faTimes~represent the functional correctness of the code.}
    \label{tab:comparison_result}
    \begin{tabular}{cccccc|cccc}
    \toprule
 & &  \multicolumn{4}{c}{\textbf{BigCodeBench}} &  \multicolumn{4}{c}{\textbf{SecCodePLT+}}\\ \cmidrule(lr){3-6} \cmidrule(lr){7-10}
 Model & Method & \faBug\faCheck{} to \faShield\faTimes &  \faShield\faCheck{} to \faShield\faTimes &  \faShield\faCheck{} to \faBug\faCheck &  \faShield{} to \faBug & \faBug\faCheck{} to \faShield\faTimes &  \faShield\faCheck{} to \faShield\faTimes &  \faShield\faCheck{} to \faBug\faCheck &  \faShield{} to \faBug \\
\hline \hline
\multirow{4}{*}{\textbf{Codellama}} & SVEN & 18.81\% & 69.00\% & 0.00\% & 3.15\% & 43.58\% & 76.74\% & 0.00\% & 4.08\%\\
 &  SafeCoder & 22.77\% & 73.00\% & 0.00\% & 3.09\% & 48.71\% & 62.79\% & 0.00\% & 5.37\%\\
 &  CodeGuard+  & 11.88\% & 33.55\% & 1.34\% & 4.07\% & 12.82\% & 39.53\% & 2.32\% & 6.42\%\\
 & PromSec  & 12.87\% & 55.70\% & 0.67\% & 1.82\% & 5.10\% & 46.51\% & 11.62\% & 12.85\%\\ \hline
 \multirow{4}{*}{\textbf{Qwen2.5-Coder}}& SVEN  & 2.59\% & 33.87\% & 0.32\% & 1.74\% & 6.12\% & 23.68\% & 7.89\% & 6.32\%\\
 &  SafeCoder & 1.03\% & 20.64\% & 0.32\% & 3.19\% & 12.24\% & 42.10\% & 5.26\% & 5.43\%\\
 &  CodeGuard+ & 1.03\% & 16.77\% & 0.32\% & 1.88\% & 2.04\% & 28.94\% & 10.52\% & 7.07\%\\
 &  PromSec & 35.75\% & 77.09\% & 0.96\% & 2.03\% & 8.16\% & 50.00\% & 5.26\% & 9.22\%\\ \hline
 \multirow{4}{*}{\textbf{DeepSeek-Coder}}& SVEN  & 47.26\% & 59.45\% & 1.35\% & 1.43\% & 10.00\% & 36.58\% & 4.87\% & 5.53\%\\
 &  SafeCoder & 13.69\% & 28.37\% & 1.35\% & 2.00\% & 4.00\% & 31.70\% & 4.87\% & 6.68\%\\
 &  CodeGuard+ & 21.23\% & 39.64\% & 0.45\% & 1.57\% & 2.00\% & 26.82\% & 2.43\% & 5.76\%\\
 & PromSec  & 43.15\% & 69.81\% & 0.00\% & 0.42\% & 0.00\% & 34.14\% & 24.39\% & 14.99\%\\ \hline
 \multirow{4}{*}{\textbf{Mistral}}&  SVEN & 8.33\% & 72.46\% & 0.00\% & 11.77\% & 42.30\% & 66.66\% & 9.09\% & 7.81\%\\
 & SafeCoder  & 13.88\% & 84.05\% & 0.00\% & 11.77\% & 3.84\% & 27.27\% & 6.06\% & 7.44\%\\
 & CodeGuard+  & 0.00\% & 8.69\% & 0.00\% & 12.76\% & 11.53\% & 45.45\% & 3.03\% & 6.71\%\\
 & PromSec  & 30.55\% & 62.31\% & 0.00\% & 12.39\% & 15.38\% & 27.27\% & 21.21\% & 18.68\%\\ \hline
 \multirow{3}{*}{\textbf{StarCoder}}& SVEN  & 40.00\% & 68.42\% & 0.00\% & 5.15\% & 0.00\% & 21.33\% & 1.33\% & 0.98\%\\
 & SafeCoder  & 0.00\% & 52.63\% & 0.00\% & 0.00\% & 0.00\% & 21.33\% & 0.00\% & 3.67\%\\
 & CodeGuard+  & 80.00\% & 57.89\% & 0.00\% & 0.73\% & 0.00\% & 14.66\% & 0.00\% & 0.71\%\\ \hline
 \multirow{1}{*}{\textbf{GPT-3.5-Turbo}}& PromSec & 20.28\% & 31.33\% & 1.63\% & 1.88\% & 3.12\% & 24.32\% & 2.70\% & 7.89\%\\ \hline
 \multirow{1}{*}{\textbf{GPT-4o}}& PromSec & 8.10\% & 16.74\% & 0.46\% & 0.44\% & 3.50\% & 29.16\% & 2.08\% & 6.02\%\\ \hline
 \multicolumn{2}{c}{\textbf{Average}} & 9.97\% & 42.04\% & 0.62\% & 4.11\% & 11.75\% & 35.04\% & 5.15\% & 6.87\% \\

\bottomrule
\end{tabular}
\end{table*}

\subsection{(RQ3) On the Performance Disparity of Functionality}
\label{subsec:rq3}

In RQ2, we observe the degradation in functionality of generated code by existing techniques. We further analyze the results to better understand this phenomenon. Specifically, we collect all tasks where the generated code by the base models is functionally correct and inspect the outcomes when applying the secure code generation techniques.
\autoref{tab:comparison_result} presents the results on the two datasets. For each dataset, the first two columns show the results for the aforementioned cases.
\begin{itemize}[topsep=0pt, leftmargin=10pt]
    \item The column ``\faBug\faCheck{} to \faShield\faTimes'' denotes cases where insecure but functional code by the base model becomes secure but non-functional after applying existing techniques.

    \item The column ``\faShield\faCheck{} to \faShield\faTimes'' represents cases where secure and functional code becomes secure but non-functional after applying existing techniques.
\end{itemize}
We observe that around 10\% to 30\% of tasks fall under the ``\faBug\faCheck{} to \faShield\faTimes'' category.
Notably, CodeGuard+ with StarCoder on BigCodeBench leads to 80\% of tasks becoming non-functional after improving security. \delete{The percentage is generally lower on the SecCodePLT+ dataset; however, for a few cases, such as SafeCoder with CodeLlama and SVEN with Mistral, it remains above 40\%.}\revise{The percentages are generally similar between BigCodeBench (Avg. 9.97\%) and SecCodePLT+ (Avg. 11.75\%), with no statistically significant difference ($p > 0.05$).}
For the ``\faShield\faCheck{} to \faShield\faTimes'' category, the percentages are even higher, with most cases ranging from 20\% to 60\% on BigCodeBench and 20\% to 40\% on SecCodePLT+.
This indicates that while existing techniques are effective in fixing vulnerabilities, they tend to compromise the functionality of generated code for samples that were previously correct. This explains the functionality degradation we observed earlier.

\revise{In summary, SVEN, SafeCoder, CodeGuard+, and PromSec can cause originally functional code to become non-functional in up to 76.74\%, 84.05\%, 80\%, and 77.09\% of cases, respectively.
We perform a two-way ANOVA test to assess whether the secure code generation method has a statistically significant effect relative to the base model. 
The results show that, for all secure code generation methods across both datasets, the p-value is less than 0.001, meaning that existing methods can significantly degrade the performance of the base model.}\looseness=-1

\begin{table}
    \centering
    \scriptsize
    \caption{Case study on the code generated by existing techniques using DeepSeek-Coder on BigCodeBench for the ``\faBug\faCheck{} to \faShield\faTimes'' category.
    The value in parentheses denotes the number of cases for each technique in this category.
    ``NFI'' represents cases where the generated code did not follow the instruction. ``FN'' indicates cases where the code snippet contains a vulnerability that was missed by the static analyzers.}
    \label{tab:case_study_insecure-fun_to_secure-non-func}
    \begin{tabular}{cccccc}
    \toprule
         & Removed Code & Junk Code & NFI & FN & Other\\ \hline\hline
      SVEN (69)   & 75.36\% & 8.7\% & 0\% & 2.89\% & 13.04\%\\
      SafeCoder (20)   & 60\% & 0\% & 0\% & 25\% & 15\%\\
      CodeGuard+ (31)   & 67.74\% & 6.45\% & 12.9\% & 3.23\% & 9.68\%\\
      PromSec (63)   & 92.06\% & 0\% & 7.94\% & 0\% & 0\%\\
      \bottomrule
    \end{tabular}
    
\end{table}

\begin{figure*}
    \centering
    \includegraphics[width=0.7\textwidth]{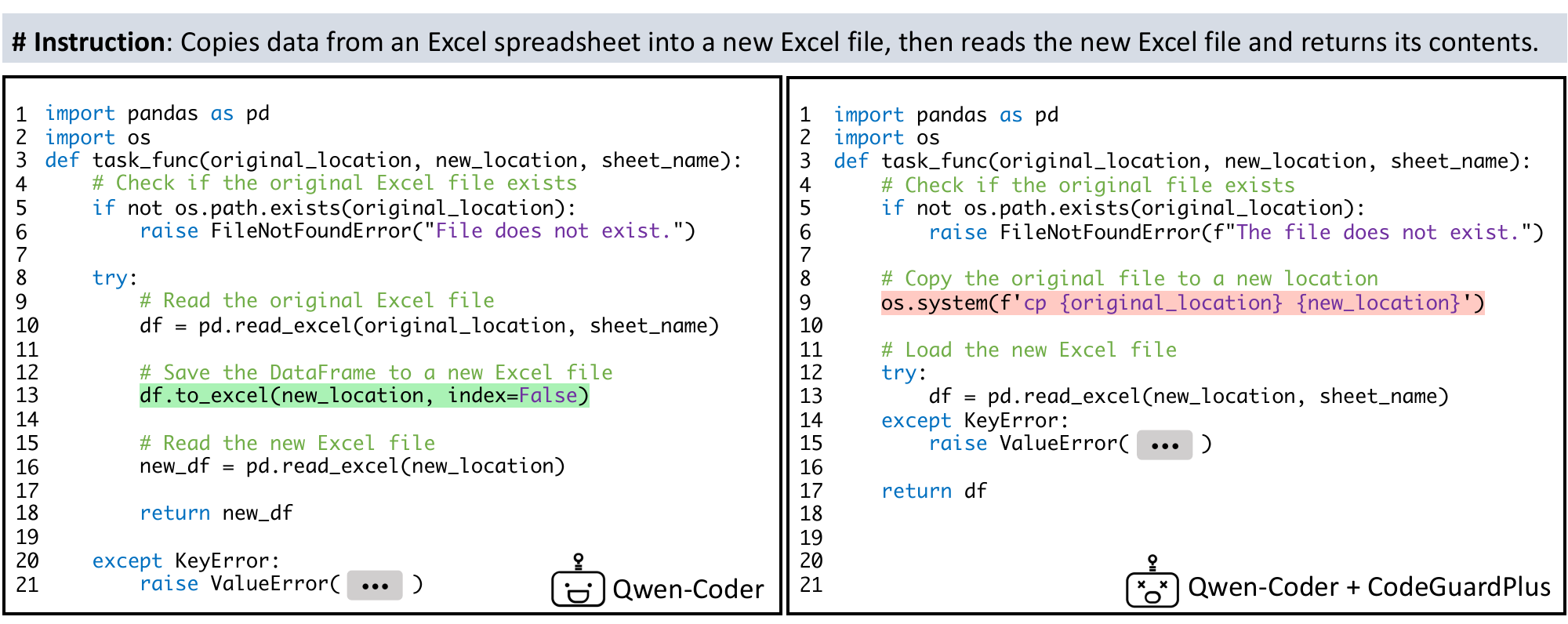}
    \caption{Generated code by Qwen before (left) and after (right) applying CodeGuard+. The code on the left is both functional and secure. However, after applying CodeGuard+, a CWE-78 vulnerability appears at line \code{9} in the code on the right. The differences are depicted in \textcolor{green}{green} and \textcolor{red}{red}, representing secure and insecure code, respectively.}
    \label{fig:case_4_safe_to_unsafe}
\end{figure*}

\begin{finding}
    \textit{\textbf{\emph{Finding 7:}} Existing secure code generation techniques negatively impact the base model, transforming originally functional code into non-functional code.}
\end{finding}

To better understand the issue, we manually inspect the cases in the ``\faBug\faCheck{} to \faShield\faTimes'' category. We use DeepSeek-Coder on BigCodeBench as an example.
\autoref{tab:case_study_insecure-fun_to_secure-non-func} presents the results for the four existing techniques. We classify the cases into five categories.
\revise{These categories are based on our manual inspection of all cases, with discrepancies resolved through majority voting among the authors.}
\begin{itemize}[topsep=0pt, leftmargin=10pt]
    \item \textbf{Removed Code:} This category includes cases where techniques simply remove vulnerability-related code to improve security, as shown in the example in \autoref{sec:motivation}. We observe a large number of cases in this category, with PromSec removing insecure code in 92.06\% of the samples.

    \item \textbf{Junk Code:} The generated output is nonsensical, such as the repeated word ``task task task.'' SVEN and CodeGuard+ show around 6\%-8\% of cases in this category.
    
    \item \textbf{Not Following Instruction (NFI):} In these cases, the LLM generates code completely irrelevant to the task. CodeGuard+ and PromSec have a non-trivial percentage of cases in this category.
    
    \item \textbf{False Negatives (FN):} These are cases where static analyzers fail to detect vulnerabilities, which was discussed in RQ1.

    \item \textbf{Other:} In a few cases, the generated code is missing necessary package imports or is otherwise incomplete, leading to functional incorrectness.
\end{itemize}

\begin{finding}
    \textit{\textbf{\emph{Finding 8:}} The security improvements achieved by existing techniques come at the cost of sacrificing the functionality of the generated code, primarily by removing vulnerability-related code and generating irrelevant or garbage output.}
\end{finding}

\subsection{(RQ4) On the Performance Disparity of Security}
\label{subsec:rq4}

While existing secure code generation techniques were designed to improve the security of LLM-generated code, they may inadvertently degrade security for certain tasks. We have already observed the performance reduction in \autoref{fig:sec_and_func_res_bigcode} and \autoref{fig:sec_and_func_res_secplt}. Here, we specifically analyze the scenario where the code generated by the base model was originally secure, but vulnerabilities are detected after applying existing techniques. We refer to this as the \textit{non-monotonic security improvement} of these methods.

\autoref{tab:comparison_result} presents the results for the scenario described above. Specifically, we consider two cases:
\begin{itemize}[topsep=0pt, leftmargin=10pt]
    \item The column ``\faShield\faCheck{} to \faBug\faCheck'' denotes cases where secure and functional code by the base model becomes insecure but remains functional after applying existing techniques.

    \item The column ``\faShield{} to \faBug'' represents cases where secure code becomes insecure after applying existing techniques, regardless of its functionality.
\end{itemize}
Observe that for the first case ``\faShield\faCheck{} to \faBug\faCheck'', although it rarely happens, around 1\% of samples can still occur, such as CodeGuard+ with Codellama and PromSec with GPT-3.5 on BigCodeBench.
The percentage is much higher in SecCodePLT+. In fact, PromSec with DeepSeek-Coder exhibits 24.39\% of samples becoming insecure after applying the technique.
When functionality is not considered, more instances of secure code becoming insecure are observed after applying existing techniques, as shown in the column ``\faShield{} to \faBug''.
The average percentages are 4.11\% for BigCodeBench and 6.87\% for SecCodePLT+, respectively.

\begin{finding}
    \textit{\textbf{\emph{Finding 9:}} The security improvement provided by existing techniques is not monotonic; they may introduce vulnerabilities into code that was previously secure.}
\end{finding}

\autoref{fig:case_4_safe_to_unsafe} shows an example where the task is to copy data from an Excel spreadsheet into a new Excel file and then read and return the contents of the new file.
The base model, Qwen2.5-Coder-7B-Instruct, can generate a secure code snippet as shown on the left.
However, after applying CodeGuard+\cite{fu2024constrained}, the generated code on the right introduces a vulnerability at line \code{9}. It uses the system call \code{os.system()} to execute the copy command, which is passed as a string. If the \code{original\_location} or \code{new\_location} variables contain malicious input, this could lead to command injection. This vulnerability corresponds to CWE-78~\cite{CWE-78}.

\section{\revise{Limitations and }Threats to Validity}

The \textit{internal} threat to validity lies in \delete{potential mistakes during manual inspection. Specifically, we may misidentify a vulnerability reported by different scanners. To mitigate this threat, we ensure that each vulnerability is examined by at least two authors. A third author will chime in to resolve any disagreements.}\revise{the limitations of static analyzers, which may produce false negatives, as discussed in~\autoref{sec: problems}. To mitigate this, we use three different static analyzers and take the union of all detected CVEs, which helps reduce false negatives.}

The \textit{external} threat to validity primarily lies in the subjects used in our study. The code generation tasks we examine may not be representative. We mitigate this risk by using two recent large datasets covering over 2,000 tasks. Since these datasets focus mainly on Python, our findings may not generalize to other programming languages. However, most existing techniques were originally evaluated using Python. Our study re-evaluates these techniques.
For the SecCodePLT dataset, we generate unit tests using an LLM. While the test cases may be limited, we address this by manually inspecting the generated inputs in conjunction with the ground truth code and setup files. We also manually calibrate problematic unit tests. \revise{Although we did not manually verify edge case coverage, adding more cases would likely lower the passing rate, which supports our conclusion that existing methods are limited.}
Additionally, the LLMs and vulnerability scanners used in this study may not be fully \delete{inclusive}\revise{representative}, especially given the rapid development of LLMs. However, we argue that the general observations regarding existing secure code generation techniques will still hold, as the issues stem from their technical design rather than the specific LLMs or tools used.\looseness=-1

\revise{The \textit{construct} threat lies in determining a code snippet as vulnerable during manual inspection. Specifically, we may misidentify a vulnerability reported by different scanners. To mitigate this threat, we ensure that each vulnerability is examined by at least two authors. A third author will chime in to resolve any disagreements.}

\smallskip\noindent
\revise{\textbf{Other Limitations.}
In our study, we do not include the HumanEval dataset as it was primarily designed to measure functionality. It consists of relatively simple tasks that lack the complexity required to trigger security issues in generated code. 
Additionally, the samples used in our evaluation might have been included in the training data of LLMs, potentially causing data contamination. Our employed two datasets are generally released after the evaluated models. Although this does not guarantee no data contamination, our results show existing works are still limited in generating secure and functional code, even if the test data may have been trained on, which further confirms the need for better designs.

Due to resource limits, we did not repeat every experiment: a full run costs 430 GPU-hours ($\approx$ 18 days) on a single NVIDIA A100, and larger models (e.g., DeepSeek) require more. As a stability check, we repeated Qwen three times with consistent results, totaling 1,290 GPU-hours ($\approx$ 54 days) on one A100.
}

\revise{\section{Discussion and Future Work}
Our study highlights key problems in current evaluation schemes: (1) relying on a single static analyzer for vulnerability detection, and (2) using separate datasets for evaluating security and functionality. Using inappropriate metrics to assess a technique’s performance could potentially mislead research progress.
To address these limitations, our study introduces a more rigorous framework that uses multiple vulnerability detectors and evaluates both security and functionality on the same set of generated code. \autoref{subsec:rq3} and \autoref{subsec:rq4} provide analysis of failure cases, offering insights into which aspects of existing methods require improvement. These contributions are crucial for establishing a proper evaluation standard.

Future research on secure code generation shall follow our evaluation framework by evaluating security and functionality jointly, such that the performance of proposed techniques is rigorously assessed and comparable with one another. Our enhanced SecCodePLT+ dataset, equipped with unit tests, can serve as a standard benchmark for comprehensively assessing the performance. Additionally, we have categorized various failure cases of existing techniques, such as removing vulnerability-related code to improve security. Future efforts shall design new approaches that can address these problems and maintain the functionality of generated code while enhancing its security. As we mainly leverage vulnerability detectors for assessing the security of generated code -- which themselves may produce false reports -- an important future direction is to develop a reliable and theoretically guaranteed vulnerability detection method.}
\section{Related Work}

\paratight{LLM for (Secure) Code Generation.}
Several Code LLMs have been specifically trained for code generation tasks using code datasets, such as CodeGen~\cite{nijkamp2023codegen}, InCoder~\cite{fried2022incoder}, SantaCoder~\cite{allal2023santacoder}, CodeLlama~\cite{roziere2023codellama}, Qwen-Coder~\cite{qwen2.5coder}, DeepSeek-Coder~\cite{deepseekcoderv2}, CodeStral~\cite{codestral}, and StarCoder~\cite{li2023starcoder}. 
Moreover, various studies have focused on enhancing the performance of Code LLMs by optimizing prompts~\cite{jiang2024self, li2024acecoder, li2025structured}. 
Additionally, some studies use LLMs to synthesize test cases and leverage the augmented datasets to improve the performance of LLM~\cite{chen2023codet, huang2023agentcoder, zeng2025acecoder}. 
Other works aiming to improve the security of LLM-generated code include SVEN~\cite{he2023large}, SafeCoder~\cite{he2024instruction}, CodeGuard+~\cite{fu2024constrained}, and PromSec~\cite{nazzal2024promsec}. We elaborate on the details of these methods in \autoref{subsec:setup}. However, all existing code generation studies have either primarily evaluated the functionality of LLM-generated code or assessed functionality and security separately.
Other applications of LLMs include program repair~\cite{xia2023automated, wei2023copiloting, zhao2024repair}, code analysis~\cite{nam2024using, fang2024large, zhang2024detecting}, and unit test generation~\cite{yuan2024evaluating, chen2024chatunitest, alshahwan2024automated}.

\paratight{LLM for Vulnerability Detection.}
Another line of research focuses on leveraging LLMs to improve code security. Several studies have utilized LLMs for vulnerability detection~\cite{ullah2024llms, ding2024vulnerability,zhou2024large,li2024llm, khare2025understanding}. 
\citet{khare2025understanding} examined LLMs' zero-shot capability in detecting vulnerabilities , and their findings show that LLMs outperform CodeQL in detecting certain vulnerabilities, such as CWE-22 and CWE-78~\cite{khare2025understanding}. 
Other works also demonstrated the LLM's promising capability in vulnerability detection~\cite{zhou2024large, sultana2024code, shestov2024finetuning}. 
In addition to leveraging LLMs for vulnerability detection, several studies have also explored their use in vulnerability repair~\cite{pearce2023examining, islam2024codesecurityvulnerabilityrepair, 10.1145/3597503.3639222}. 

\paratight{Evaluation of LLM-generated Code.}
Several benchmarks have been proposed for evaluating the functional correctness of LLM-generated code, including LiveCodeBench~\cite{jain2024livecodebenchholisticcontaminationfree}, BigCodeBench~\cite{zhuo2024bigcodebench}, HumanEval~\cite{chen2021evaluatinglargelanguagemodels}, MBPP~\cite{DBLP:journals/corr/abs-2108-07732}, and SWE-Arena~\cite{swe-arena2024}. Other benchmarks that consider both functionality and security, such as SecCodePLT~\cite{yang2024seccodeplt}, CodeGuard+~\cite{fu2024constrained}, and CWEval~\cite{peng2025cweval}, are either limited by dataset size~\cite{fu2024constrained, peng2025cweval} or lack test cases for nearly half of the samples~\cite{yang2024seccodeplt}.
Additionally, there are datasets specifically designed for code security evaluation, including SecurityEval~\cite{siddiq2022securityeval}, LLMSecEval~\cite{10174231}, CyberSecEval~\cite{bhatt2023purple}, and Sallam~\cite{10.1145/3691621.3694934}. However, these datasets do not provide unit tests for functional evaluation.
We employ BigCodeBench and SecCodePLT in our study and extend SecCodePLT by augmenting its unit test cases.

Other than using benchmarks for evaluation, several tools and metrics exist for assessing the functionality or security of code. \citet{ren2020codebleumethodautomaticevaluation} proposed CodeBLEU, which evaluates the accuracy of generated code compared to the ground truth by considering n-gram matches, AST matches, and data-flow matches. \citet{le2024indict} prompted LLMs to judge the security and helpfulness of code, while \citet{tong-zhang-2024-codejudge} examined different prompting techniques for evaluating the semantic correctness of LLM-generated code. \citet{pearce2022asleep, siddiq2022securityeval} analyzed the safety rate of the generated code using CodeQL. 
\citet{bhatt2023purple} proposed an insecure code detector, which leverages a static analyzer and aims to detect insecure coding practices rather than specific vulnerabilities. Finally, \citet{wang2023enhancinglargelanguagemodels} conducted a study on LLMs' capabilities in secure code generation, vulnerability classification, vulnerability repair, and vulnerability explanation.
However, its goal is not to evaluate existing secure code generation techniques.
\section{Conclusion}
We conduct a comprehensive study of four secure code generation techniques across two benchmarks.
Our study results suggest that future work should employ more than one vulnerability scanner for security evaluation, as different scanners have varying strengths.
We also show that existing techniques have limited effectiveness in enhancing the security of LLM-generated code when considering functionality simultaneously. These techniques tend to sacrifice functionality to achieve a higher security score, leading to non-usable generated code.
Our study underscores the importance of evaluating both the security and functionality of LLM-generated code simultaneously and provides guidelines for future research.

\begin{acks}
This work was supported by NVIDIA Academic Grant Award, National Science Foundation awards \#2340198, \#2319880, and \#2213727, and Cisco University Research Program Fund \#71858473. 
\end{acks}

\bibliographystyle{ACM-Reference-Format}
\bibliography{reference}

\end{document}